\documentclass[fleqn, useAMS,usenatbib]{mn2e}

\usepackage{ amssymb, graphicx}
\usepackage{ upgreek, amsmath}

\newcommand{\e}[1]{\times 10^{#1}}

\newcommand{\wl}{$\lambda$}
\newcommand{\msun}{M$_\odot$}
\newcommand{\wll}{$\lambda \lambda$}

\setcitestyle{round,aysep={},yysep={,}}
\graphicspath{ {/Users/anders/projects/finished/Jerkstrand2014_SN2012aw/figs/} }

\title[SN 2012aw]{The nebular spectra of SN 2012aw and constraints on stellar nucleosynthesis from oxygen emission lines }
\author[Anders Jerkstrand]{A. Jerkstrand$^{1}$\thanks{E-mail:a.jerkstrand@qub.ac.uk},
S. J. Smartt$^{1}$, M. Fraser$^{1,3}$, C. Fransson$^{2}$, J. Sollerman$^{2}$, F. Taddia$^{2}$, 
\newauthor R. Kotak$^1$\\
$^1$Astrophysics Research Centre, School of Mathematics and Physics, Queen's University Belfast, Belfast BT7 1NN, UK
\\
$^2$The Oskar Klein Centre, Department of Astronomy, Stockholm University, Albanova, 10691 Stockholm, Sweden\\
$^3$Institute of Astronomy, University of Cambridge, Madingley Road, Cambridge CB3 0HA, UK
}
\begin{document}

\date{Accepted Jan 30 2014--- Received --- in original form ---}

\pagerange{\pageref{firstpage}--\pageref{lastpage}} \pubyear{2002}

\maketitle

\label{firstpage}

\begin{abstract}
We present nebular phase optical and near-infrared spectroscopy of the Type IIP supernova SN 2012aw combined with NLTE
radiative transfer  calculations applied to ejecta from stellar evolution/explosion models. Our spectral synthesis models generally
show  good agreement with the ejecta from a $M_{\rm ZAMS}=15$ \msun\ progenitor star. The emission lines of oxygen, sodium, and magnesium are all consistent with the nucleosynthesis in a progenitor in the $14-18$ \msun\ range. We also demonstrate how the evolution of the  oxygen cooling lines of [O I] \wl 5577, [O I] \wl6300, and [O I] \wl6364 can be used to constrain the mass of oxygen in the non-molecularly cooled ashes to $< 1$ \msun, independent of the mixing in the ejecta. This constraint implies that any progenitor model of initial mass greater than 20 \msun\ would be difficult to reconcile with the observed line strengths.  A stellar progenitor of around $M_{\rm ZAMS}=15$ \msun\  can consistently explain the directly measured luminosity of the progenitor star, the observed nebular spectra, and the inferred pre-supernova mass-loss rate. 
We conclude that there is still no convincing example of a Type IIP supernova showing the nucleosynthesis products expected from a $M_{\rm ZAMS} > 20$ \msun\ progenitor.

\end{abstract}

\section{Introduction}
An understanding of the origin of the elements in the universe requires an understanding of the nucleosynthesis in Type IIP supernovae (SNe) and their progenitors, which make up 40\% of all SN explosions in the local universe \citep{Li2011}. These are the explosions of stars that have retained most of their hydrogen envelopes throughout their evolution, which results in a $\sim$100 day hydrogen recombination plateau in the light curve. After this follows the nebular phase, when the inner regions of the ejecta become visible. 

 With recently produced SN ejecta grids from a range of progenitor masses, improvements in atomic data, and the development of NLTE (Non-Local Thermodynamic Equilibrium) radiative transfer codes \citep[e.g.,][]{Dessart2011,Jerkstrand2011,Maurer2011}, self-consistently calculated synthetic spectra for these kind of objects are now achievable. 
These models can be used to diagnose the nucleosynthesis in individual events and improve our understanding of the role of Type IIP SNe for the cosmic production of abundant intermediate-mass elements such as oxygen.

Nucleosynthesis analysis can be performed in the nebular phase of the SN evolution, when the photosphere has receded to reveal the products of hydrostatic and explosive burning. Since the nucleosynthesis depends strongly on the main-sequence mass of the star \citep[e.g.][]{Woosley1995}, comparing the strengths of observed nebular lines to models allows an estimate of the main-sequence mass. 

In the cases where pre-explosion images of the progenitor are available, another powerful technique for estimating the progenitor mass is by comparing the luminosity of the progenitor with stellar evolutionary models  \citep[e.g.][]{Smartt2009a}. Combining such progenitor analysis with nebular phase spectral modelling offers a way of tightly constraining stellar evolution models and nucleosynthesis. A third approach to analyse the link between progenitors and SN observables is through hydrodynamical modeling of the diffusion-phase light curve \citep[see e.g.][]{Utrobin2008, Utrobin2009, Bersten2011, Pumo2011}.

In \citet[][J12 hereafter]{Jerkstrand2012}, we developed models for the spectra of Type IIP SNe from the onset of the nebular phase ($\sim140$ days) up to 700 days, to use for spectral analysis. These models solve for the non-thermal energy deposition channels, the statistical and thermal equilibrium for each layer of nuclear burning ashes, and the internal radiation field. The calculations are currently applied to SN ejecta from stars evolved and exploded with KEPLER \citep{Woosley2007}.

Here, we apply these models to optical and near-infrared observations of SN 2012aw, a nearby Type IIP SN \citep{Bose2013, dallOra2013} with a potentially luminous and massive progenitor directly identified in pre-explosion images \citep{Fraser2012, vandyk2012,Kochanek2012}. The inferred luminosity of the progenitor was disputed in these separate analyses (see Sect. \ref{sec:discussion}) but is potentially 
higher than most previous red supergiant progenitor stars and a candidate to be the first star with $M_{\rm ZAMS} > 20$ \msun\ (from progenitor analysis) seen to explode as a Type IIP SN. The lack of any such stars in the sample of detections and upper limits obtained so far has been termed the ``red supergiant problem'' \citep{Smartt2009b}, and is currently challenging our understanding of the final outcomes of $M_{\rm ZAMS}=20-30$ \msun\ stars, and their role in galactic nucleosynthesis.
The strong theoretical dependency of the oxygen yield with progenitor mass makes SN 2012aw an interesting candidate to analyze for signs of the high mass of synthesised oxygen expected from a $M_{\rm ZAMS} > 20$ \msun\ progenitor.

\section{Observational data}
\label{sec:obs}
We obtained four optical ($250-451$ days) and one near-infrared (306 days) spectra of SN 2012aw using the William Herschel Telescope (WHT) and the Nordic Optical Telescope (NOT), as detailed in Table \ref{tab:log}. The NIR spectrum at 306 days post-explosion is among the latest near-infrared spectra taken for a Type IIP SN. The optical spectra were taken with ISIS (WHT) and ALFOSC (NOT), while the near-infrared (NIR) WHT spectrum was taken with LIRIS.  All ISIS data were reduced within {\sc iraf} using standard techniques. The spectra were bias-subtracted and flat-fielded using a normalised flat field from an internal lamp. Cosmic rays were removed in the 2D spectral images using {\sc lacosmic} \citep{vandokkum2001}, before individual exposures were combined, and the 1D spectrum was optimally extracted. Arc spectra from CuNe and CuAr lamps were used to wavelength calibrate the extracted spectra, while flux calibration was performed using observations of spectrophotometric standard stars taken with the same instrumental configuration on the same night. No attempt was made to correct for telluric absorptions. 
The NOT spectra were reduced in a similar manner,  including corrections for second order contamination \citep{Stanishev2007}.  

To calibrate the spectra to photometry, we scaled each spectrum with a constant to minimize the RMS difference between synthetic photometry calculated from the spectra and observed Johnson-Cousins $BVRI$ photometry. For the day 250 spectrum we used the day 248 photometry of \citet{Bose2013} and for the day 332 spectrum we used the day 333 photometry of \citet{dallOra2013}. For the day 369 and day 451 spectra, we employed an extrapolation of the \citet{dallOra2013} day 333 photometry assuming decline rates based on the \citet{Sahu2006} observations of SN 2004et (which had a spectral evolution similar to SN 2012aw) between $336-372$ days for the day 369 spectrum and between $336-412$ days for the 451 day spectrum. We obtain RMS errors of 4.6\% (250d), 8.0\% (332d), 7.3\% (369d), and 11\% (451d) for the $BVRI$ scatter, which we regard as representative of the absolute flux errors.

\begin{table*}
\caption{Log of  spectroscopy of SN 2012aw.  The phase is with respect to the explosion epoch of March 16 2012 \citep{Fraser2012}. The resolution is the FWHM resolution at the midpoint of the spectral range.}
\begin{tabular}{lcccccccc}
Date			& Phase  &Instrument	&  Disperser	    & Coverage     & Exp. Time    &  Slit	& Dispersion           &	Resolution 	 	\\
                        & (days)   &               &                  & (\AA)      & (seconds)    & (\arcsec)         & (\AA/pixel) & (\AA) \\
\hline
2012-11-21	& +250	  &   WHT+ISIS & R300B/R158R  & 3300-9500    & 1800       & 2.0  & 0.86/1.8   & 8.2/15.4\\  
2013-01-16	& +306	  & WHT+LIRIS & zJ+HK         & 8870-24000  & 2000/1600  & 1.0  & 6.1/10    	&  14/30   	\\ 

2013-02-11	& +332	      & WHT+ISIS & R300B/R158R    & 3300-9500   &    1800          & 1.5 & 0.86/1.8	 & 6.2/11.6\\	 
2013-03-20         & +369       & NOT+ALFOSC & Grism 4                & 3330-9140  &   5400  & 1.0   &  2.95   & 16 \\ 
2013-06-10	& +451	      & WHT+ISIS & R300B/R158R    &  3300-9500  &      2700    & 1.0  & 0.86/1.8	&  4.1/7.7	\\ 
\hline
\end{tabular}
\label{tab:log}
\end{table*}
For the LIRIS NIR spectrum, we alternated the target between two positions in
the LIRIS slit. By subtracting pairs of spectra at different slit
positions from each other, the sky background and detector bias were removed. LIRIS suffers from a problem where pixels may be shifted during
readout, so we first ``descrambled'' the raw data to correct for this, using the {\sc lcpixmap} task with the {\sc liris} data reduction package within {\sc iraf} \footnote{http://www.ing.iac.es/astronomy/instruments/liris/liris\_ql.html}.
The 2D spectra were then flat-fielded and shifted to spatially align the 
spectra. They were then extracted and wavelength
calibrated using Xe and Ar arc lamps. A solar analog was
observed in between the zJ and HK spectra of SN 2012aw at a similar airmass, and this was
used to correct for the telluric absorption. As there are few good
spectrophotometric standards in the NIR, we performed an approximate
flux calibration, estimated by scaling the flux of the telluric standard
to match its catalogued 2MASS {\it JHK} magnitudes. The resulting spectrum
showed a $\sim$30\% flux deficiency in the overlap region with both the day
250 and day 332 optical spectra (scaled with exponential decay factors to the same epoch). We therefore applied an additional
correction factor of 1.3 to force overlap between the NIR spectrum and the optical spectra.

To compare the observed spectra to models, we adopt a distance of 9.9 Mpc \citep{Bose2013}, an extinction of $E_{B-V}=0.074$ mag \citep{Bose2013} and the extinction law of \citet{Cardelli1989}, with $R_V=3.1$. The uncertainties in distance and extinction are small (0.1 Mpc and 0.008 mags, respectively, \citet{Bose2013}). We also adopt a recessional velocity of 778 km s$^{-1}$ \citep{Bose2013}, and an explosion epoch of March 16 2013 (MJD 56002) from  \cite{Fraser2012}. The ejected mass of $^{56}$Ni is well constrained to be $0.06\pm 0.01$ \msun\ by the evolution in the early nebular phase when gamma-ray trapping is almost complete \citep{Bose2013}. 
\section{Modelling}
We compute $M_{ZAMS}=12,\ 15\ \mbox{and}\ 19$ \msun\ models at the observational epochs of SN 2012aw, following the same method and model setup as described in J12, apart from a few minor code updates described below. The model setup procedure (as more completely described in J12) involves dividing the ejecta into several chemically distinct zones - Fe/He, Si/S, O/Si/S, O/Ne/Mg, O/C, He/C, He/N, and H, named after their dominant components. To mimic the mixing structures obtained in multidimensional explosion simulations, the metal zones (Fe/He, Si/S, O/Si/S, O/Ne/Mg, O/C) are macroscopically mixed together with parts of the He/C, He/N, and H zones (fractions 0.6, 0.6, and 0.15 respectively) in a core region between 0 and 1800 km s$^{-1}$. Each zone has an individual filling factor in this core, and is distributed over $10^3$ (identical) clumps. The model takes dust formation into account by applying a gray opacity over the core from 250 days, growing with $d \tau_{dust} = 1.8\e{-3}$ per day (a calibration to the observed dust emission of SN 2004et). Over the period covered here, this dust component has only a small effect on the optical/NIR spectrum as $\tau_{dust} < 0.18$ at all times. Outside the core reside the remaining helium layers followed by the hydrogen envelope, whose density profile is determined by the one-dimensional hydrodynamic solutions. The exact zone masses, chemical compositions, and density profiles for the 12, 15 and 19 \msun\ models are given in J12. In addition to these, we here also compute the spectral evolution of the $M_{\rm ZAMS}=25$ \msun\ model of \citet{Woosley2007}, which was set-up following the same method. Table \ref{table:25} shows its zone masses, filling factors, and chemical composition.

We make a few minor updates to the code compared to the version used in J12. First, we replace the non-relativistic Doppler formula with the relativistic one \citep[e.g.][]{Rybicki1979}
\begin{equation}
\lambda = \lambda' \gamma(V) \left(1 - \frac{V}{c}\cos{\theta}\right)
\end{equation}
where $\lambda$ and $\lambda'$ are the wavelengths in the two frames, $V$ is the relative velocity, $\theta$ is the angle (between the direction of the photon and the velocity), $c$ is the speed of light, and $\gamma=\left(1-V^2/c^2\right)^{-1/2}$. We find this correction to give small but discernible differences in the line profiles, useful for detailed analysis of dust effects etc. Second, we increase the number of photon packets in the Monte Carlo simulations to keep photon noise to a minimum. Finally, we correct a minor error in the J12 calculations where the first photoexcitation rate in each NLTE solution was accidentally set to zero. Recomputation shows that this did not have any noticeable effects on the spectra apart from a 5-10\% shift in the Na I D line luminosity.

\begin{table*}

\caption{The 25 \msun\ model. The total ejecta mass is 13.7 \msun. The core filling factors are denoted $f_{core}$. Mass fractions below $10^{-9}$ are put to zero in the model.}
\begin{tabular}{|c|c|c|c|c|c|c|c|c|}
\hline
\hline
Zone                & Fe/He       & Si/S        & O/Si/S       & O/Ne/Mg       & O/C         &   He/C       & He/N        & H \\
\hline
Mass $(M_\odot$)     & 0.064       &  0.18      & 0.62          & 2.7           & 1.4         & 1.0          & 0.35        & 7.4 \\
$f_{core}$                   &  0.15       & 0.10       & 0.028         & 0.15          & 0.077       & 0.081        & 0.027       & 0.39\\
\emph{Mass fractions:}\\
$^{56}$Ni + $^{56}$Co & 0.77        &  0.075      & $4.8\e{-7}$ & $1.3\e{-7}$  & $3.9\e{-8}$  & $4.2\e{-8}$  & $2.9\e{-8}$ & 0\\
$^{57}$Co            & 0.031       & $1.7\e{-3}$ & $4.4\e{-6}$  & $1.7\e{-7}$ & $2.1\e{-8}$   & $6.8\e{-9}$  & $1.4\e{-9}$ & 0 \\
$^{44}$Ti            & $3.4\e{-4}$ & $1.8\e{-5}$ & $1.7\e{-6}$  & 0            & 0            & 0            & 0           & 0\\
H                   & $7.5\e{-8}$  & $1.3\e{-9}$ & 0           & 0            & 0             & 0           & $6.7\e{-8}$  & 0.53\\
He                  & 0.12         & $1.1\e{-5}$ & $1.9\e{-6}$ & $1.5\e{-6}$  & $2.6\e{-5}$   & 0.96        & 0.99         & 0.45     \\
C                   & $2.7\e{-6}$  & $6.4\e{-7}$ & $8.6\e{-5}$ & $8.1\e{-3}$  & 0.20         & 0.017        & $3.2\e{-4}$  & $8.6\e{-4}$ \\
N                   & $2.8\e{-7}$  & 0           & $2.2\e{-5}$ & $4.2\e{-5}$  & $1.5\e{-5}$  & $7.7\e{-6}$  & $9.0\e{-3}$  & $5.5\e{-3}$ \\
O                   & $7.9\e{-6}$  & $6.7\e{-6}$ & 0.55        & 0.74         & 0.75         & $4.1\e{-3}$  & $1.6\e{-4}$  & $3.5\e{-3}$ \\
Ne                  & $8.6\e{-6}$  & $1.7\e{-6}$ & $1.5\e{-4}$ & 0.16         & 0.034        & 0.014        & $1.1\e{-3}$  & $1.2\e{-3}$ \\
Na                  & $4.8\e{-7}$  & $7.4\e{-7}$ & $5.5\e{-6}$ & $2.2\e{-3}$  & $2.0\e{-4}$  & $1.9\e{-4}$  & $1.8\e{-4}$  & $9.9\e{-5}$ \\
Mg                  & $1.9\e{-5}$  & $1.7\e{-4}$ & 0.014       & 0.063        & $7.1\e{-3}$  & $5.8\e{-4}$  & $5.7\e{-4}$  & $6.1\e{-4}$   \\
Al                  & $6.7\e{-6}$  & $2.2\e{-4}$ & $4.5\e{-4}$ & $6.3\e{-3}$  & $1.0\e{-4}$  & $7.5\e{-5}$  & $1.0\e{-4}$  & $6.9\e{-5}$\\
Si                  & $2.3\e{-4}$  & 0.37        & 0.26        & 0.013        & $8.9\e{-4}$  & $8.3\e{-4}$  & $8.2\e{-4}$  & $8.2\e{-4}$ \\
S                   & $1.9\e{-4}$  & 0.39        & 0.15        & $3.9\e{-4}$  & $2.3\e{-4}$  & $4.1\e{-4}$  & $4.2\e{-4}$  & $4.2\e{-4}$ \\
Ar                  & $1.5\e{-4}$  & 0.060       & 0.022       & $8.1\e{-5}$  & $8.6\e{-5}$  & $1.1\e{-4}$  & $1.1\e{-4}$  & $1.1\e{-4}$\\
Ca                  & $1.7\e{-3}$  & 0.042       & $5.0\e{-3}$ & $2.7\e{-5}$  & $2.8\e{-5}$  & $7.3\e{-5}$  & $7.4\e{-5}$  & $7.4\e{-5}$  \\
Sc                  & $2.1\e{-7}$  & $2.7\e{-7}$ & $2.7\e{-7}$ & $1.5\e{-6}$  & $7.8\e{-7}$  & $8.5\e{-8}$  & $4.5\e{-8}$  & $4.5\e{-8}$\\
Ti                  & $1.1\e{-3}$  & $5.8\e{-4}$ & $4.8\e{-5}$ & $7.9\e{-6}$  & $8.0\e{-6}$  & $3.4\e{-6}$  & $3.4\e{-6}$  & $3.4\e{-6}$ \\
V                   & $1.3\e{-5}$  & $1.5\e{-4}$ & $3.8\e{-6}$ & $5.2\e{-7}$  & $3.2\e{-7}$  & $4.9\e{-7}$  & $4.3\e{-7}$  & $4.3\e{-7}$ \\
Cr                  & $1.6\e{-3}$  & $7.5\e{-3}$ & $3.3\e{-5}$ & $1.2\e{-5}$  & $1.3\e{-5}$  & $2.0\e{-5}$  & $2.0\e{-5}$  & $2.0\e{-5}$ \\
Mn                  & $2.1\e{-6}$  & $3.3\e{-4}$ & $2.3\e{-6}$ & $2.8\e{-6}$  & $2.1\e{-6}$  & $1.8\e{-5}$  & $1.5\e{-5}$  & $1.5\e{-5}$\\
Fe                  & $9.7\e{-4}$  & 0.047       & $4.7\e{-4}$ & $5.7\e{-4}$  & $5.8\e{-4}$  & $1.4\e{-3}$  & $1.4\e{-3}$  & $1.4\e{-3}$ \\
Co                  & $3.7\e{-8}$  & $3.1\e{-9}$ & $1.7\e{-6}$ & $1.6\e{-4}$  & $1.7\e{-4}$  & $4.4\e{-6}$  & $4.0\e{-6}$  & $4.0\e{-6}$  \\
Ni                  & 0.037        & $2.7\e{-3}$ & $1.1\e{-3}$ & $6.6\e{-4}$  & $6.8\e{-4}$  & $8.2\e{-5}$  & $8.2\e{-5}$  & $8.2\e{-5}$  \\
\hline
\end{tabular}
\label{table:25}
\end{table*}

All model spectra presented in the paper have been convolved with a Gaussian with $FWHM=\lambda/500$, corresponding to the typical resolving power of the instruments.

\section{Results}
\begin{figure*}
\includegraphics[trim=10mm 0mm 18mm 0mm, clip, width=0.95\linewidth]{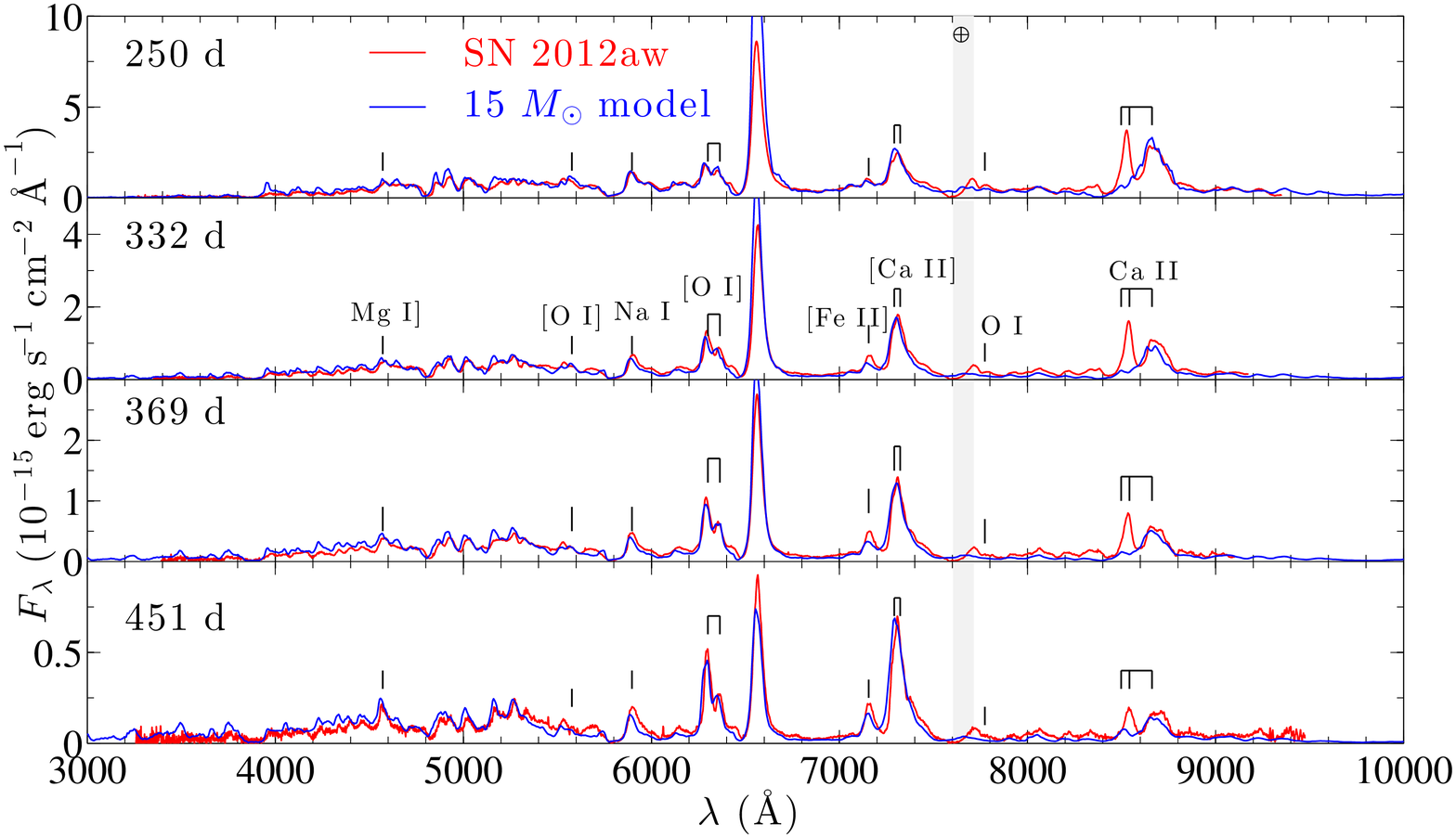}
\caption{Observed (dereddened and redshift corrected) spectra of SN 2012aw (red) compared to the 15 $M_{\odot}$ model (blue). The gray band shows the strongest telluric absorption band.}
\label{fig:optical}
\end{figure*}

\begin{figure*}
\includegraphics[trim=10mm 0mm 18mm 0mm, clip, width=0.95\linewidth]{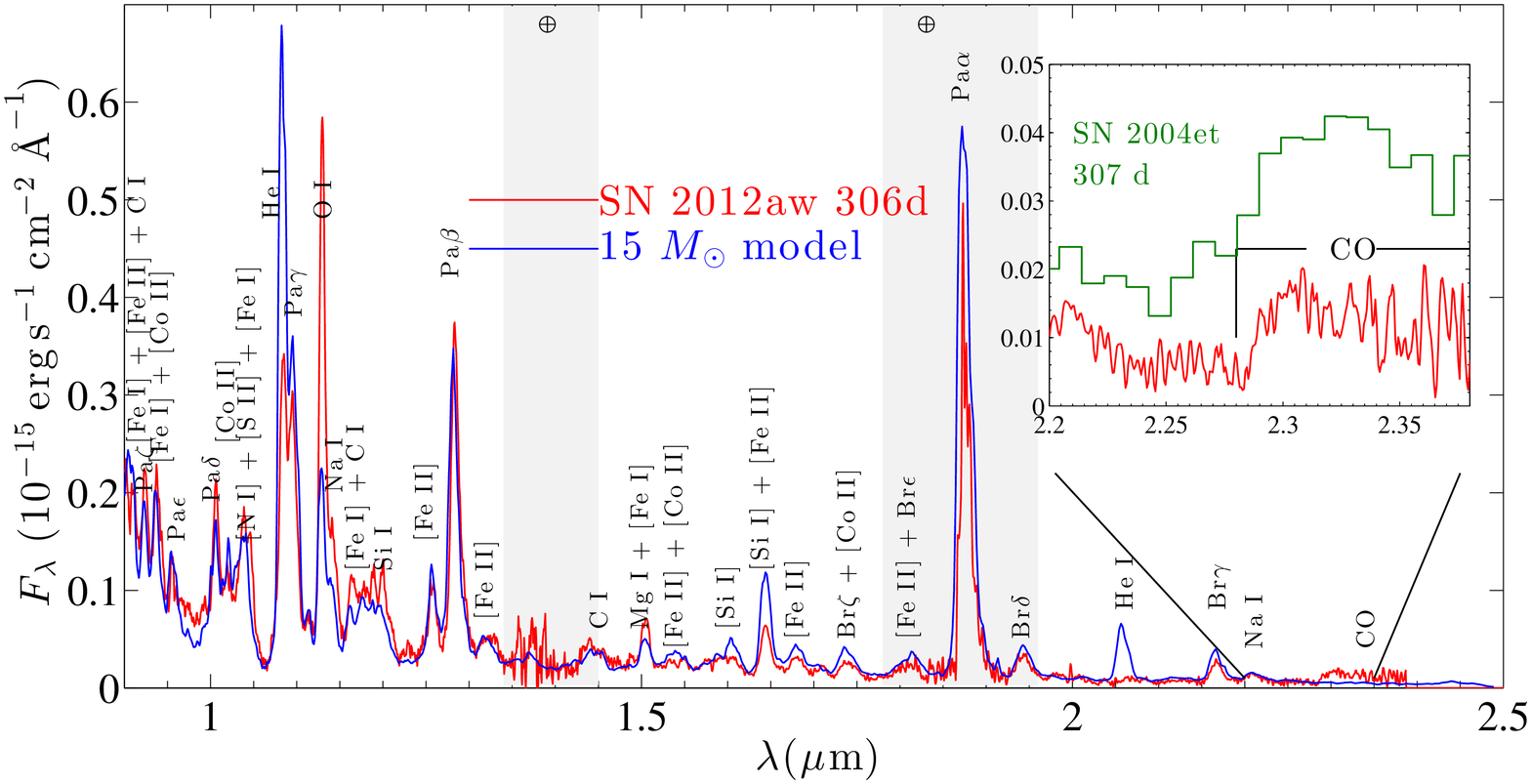}
\caption{Observed (dereddened and redshift corrected) near-infrared spectrum of SN 2012aw at 306 days (red) compared to the 15 $M_{\odot}$ model (blue). Inset is a zoom-in on the CO overtone, and a comparison with the overtone in SN 2004et at the same epoch \citep{Maguire2010}, scaled to the same distance assuming $D=5.5$ Mpc and $E_{B-V}=0.41$ mag for SN 2004et. The gray bands show the strongest telluric absorption bands.}
\label{fig:NIR}
\end{figure*}

Figs. \ref{fig:optical} and \ref{fig:NIR} show the observed (dereddened and redshift corrected) optical and near-infrared spectra of SN 2012aw, compared to the $M_{\rm ZAMS}=15\ M_{\odot}$ model computed at the corresponding epochs.

The model has an initial $^{56}$Ni mass of 0.062 \msun, and the agreement in flux levels during the early radioactive tail phase shows that SN 2012aw has a similar $^{56}$Ni mass, as also found by \citet{Bose2013}. Spectroscopically, there is satisfactory agreement throughout the evolution, showing that the ejecta model is representative of the SN 2012aw ejecta and that the model calculations capture the main physical processes important for the formation of the spectrum. The main shortcomings of the model - too strong H$\alpha$, Pa$\alpha$, He I \wl1.083 $\upmu$m, He I \wl2.058 $\upmu$m,  as well as too strong scattering in Ca II \wll8542, 8662, are the same issues that were present in the analysis of SN 2004et (J12). The incorrect Ca II NIR triplet reproduction occurs as Ca II \wl8498 and Ca II \wl8542 scatter in Ca II \wl8662 in the model, presumably due to a too high density in the envelope, or a poorly reproduced ionization balance in calcium (the relative amounts of Ca II and Ca III). The H and He line discrepancies are likely related to the mixing-scheme applied for bringing hydrogen and helium clumps into the radioactive core, but do not strongly affect our analysis of the metal emission lines below.

There is a unique emission line that provides a direct link to the nucleosynthesis - the [O I] \wll 6300, 6364 doublet. The usefulness of this line stems from several factors. Oxygen production is highly sensitive to the helium-core mass, which in turn depends on the zero-age main sequence (ZAMS) mass \citep[e.g.][]{Woosley1995, Thielemann1996}. Additionally, most of the oxygen is synthesized during hydrostatic helium and carbon burning, with little creation or destruction in the explosive process. Finally, the carbon burning ashes (the O/Ne/Mg zone) that contain most of the synthesised oxygen are devoid of much carbon and silicon, which prevents the formation of any large amounts of CO and SiO (at least for $t < 500$\ days, \citet{Sarangi2013}) that would damp out the thermal [O I] \wll 6300, 6364 emission, as happens in the helium and neon burning ashes \citep[e.g][]{Liu1995}. For SN 2004et, the [O I] \wll 6300, 6364 luminosity was about as strong as the combined molecular emission in fundamental and overtone bands of CO and SiO \citep[][J12]{Kotak2009}, showing that the bulk of the oxygen is not molecularly cooled. In SN 2012aw, the CO overtone at 2.3 $\upmu$m is similar in strength to SN 2004et (Fig. \ref{fig:NIR}), consistent with the amount of molecular cooling being similar in these two objects.

In the absence of significant molecular cooling, the [O I] \wll 6300, 6364 doublet becomes the main cooling agent of the carbon burning ashes, reemitting $50-70$\% of the deposited thermal energy in our models in the $300-700$ day interval. Since a larger mass of oxygen-rich ashes absorbs a larger amount of gamma-rays, the line is a direct tracer of the oxygen production. The analysis is aided by the oxygen density constraints set by the evolution of the [O I] \wl 6300/[O I] \wl 6364 line ratio \citep{Spyromilio1991, Li1992}, although for low-mass stars ($M_{\rm ZAMS} \lesssim 12$ \msun)\ contributions by primordial oxygen in the hydrogen envelope necessitates a multi-component analysis \citep{Maguire2012}. The density used in the model is the one derived from observations of the [O I] \wl 6300/[O I] \wl 6364 line ratio in SN 1987A. It is clear from Fig. \ref{fig:optical} that the observed line ratio in SN 2012aw evolves in a manner consistent with the model, and thus the oxygen density in SN 2012aw must be similar to the one in SN 1987A.
Also many other Type IIP SNe show a similar evolution of this line ratio, suggesting that the oxygen density shows little variation within the Type IIP class \citep{Maguire2012}.

\begin{figure}
\includegraphics[trim=0mm 10mm 17mm 15mm, clip, width=1\linewidth]{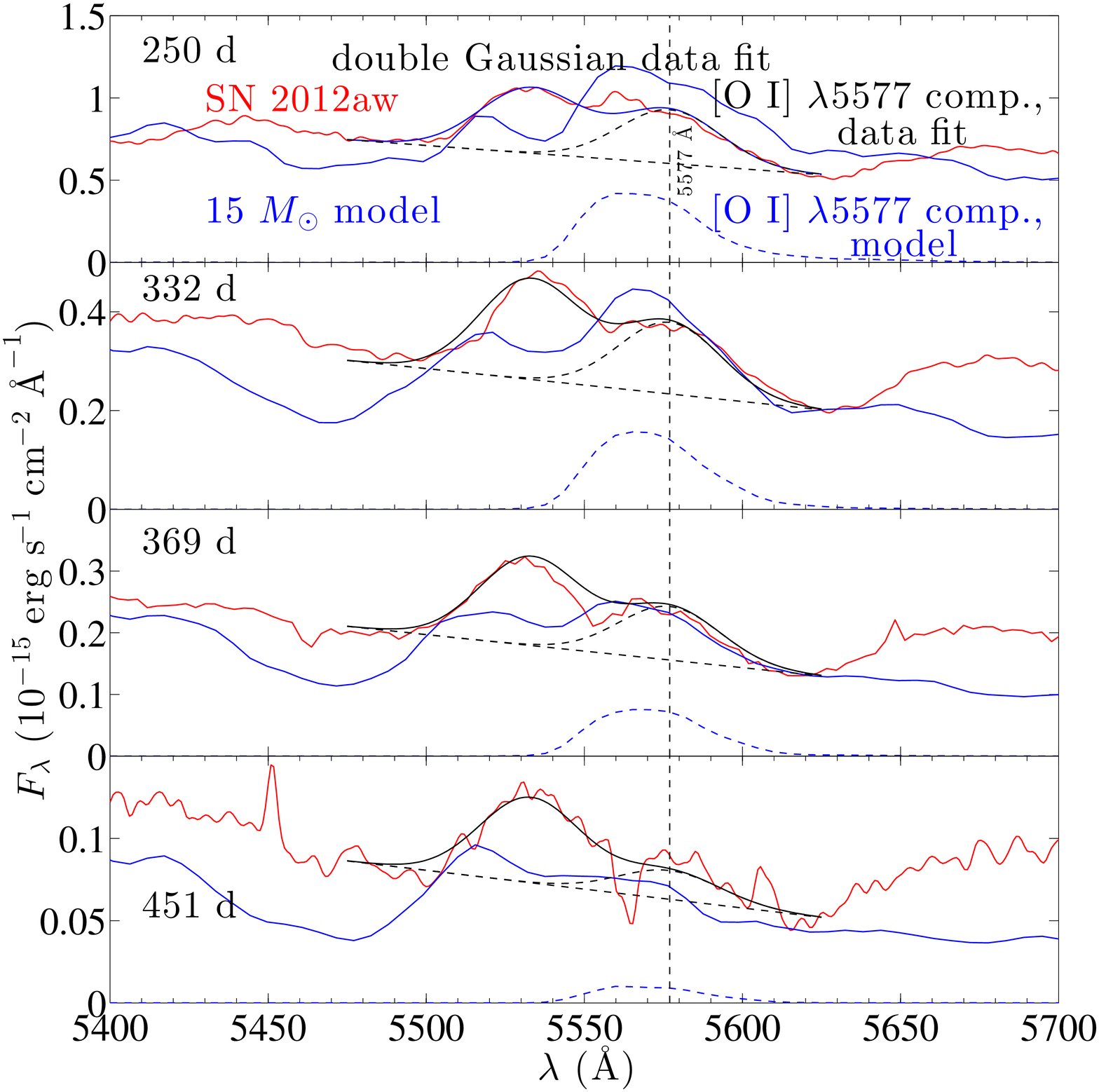}
\caption{SN 2012aw (red, solid) between 5400-5700 \AA\ (dereddened and redshift corrected), the double Gaussian fit to [Fe II] \wl 5528 and [O I] \wl 5577 (black, solid), the [O I] \wl 5577 component to the fit (black, dashed), the 15 \msun\ model (blue, solid) and the [O I] \wl 5577 component in the 15 \msun\ model (blue, dashed). The Gaussian central wavelengths are fixed at 5533 \AA\ (the 270 km s$^{-1}$ offset from 5528 \AA\ may be due to contribution by other lines or possibly by an asymmetry in the iron distribution) and 5577 \AA, and the widths are fixed at $FWHM = 40$ \AA.}
\label{fig:oi5577}
\end{figure}

Another important aspect of the 
[O I] \wll 6300, 6364 modelling is to have a representative  morphological structure for how the gamma-ray emitting $^{56}$Co clumps are hydrodynamically mixed with the oxygen clumps, since this determines the amount of gamma-ray energy deposited into the oxygen clumps. The model we use assumes a uniform distribution of $^{56}$Ni and O clumps between zero and 1800 km s$^{-1}$, a scenario that emerges by Rayleigh-Taylor mixing in the explosion \citep[e.g.][]{Kifonidis2006} and is empirically supported by the observed similarity between iron and oxygen emission line profiles (J12, see also Fig. \ref{fig:oi5577} here). A check on this mixing treatment is nevertheless desirable and the ratio of [O I] \wl 5577 to [O I] \wll 6300, 6364 can help. Whereas [O I] \wll 6300, 6364 has been widely discussed and used in previous nebular analyses, [O I] \wl 5577 has rarely been used since it is about an order of magnitude weaker than [O I] \wll 6300, 6364 and lies in a more blended spectral region. But if a luminosity can be extracted, the [O I] \wl 5577 line is useful as the [O I] \wl 5577/[O I] \wll 6300,~6364 ratio is sensitive to the temperature and therefore the gamma-ray deposition per unit mass. Both lines are excited by thermal collisions, but as they have different excitation energies their line ratio depends on the temperature (and density) \citep[see e.g.][]{Fransson1989}. 

An emission line that we identify with [O I] \wl 5577 is clearly detected in SN 2012aw (Fig. \ref{fig:oi5577}), although it is partially blended with another line that we identify with [Fe II] \wl5528. To measure the luminosity in [O I] \wl 5577 we simultaneously fit the [Fe II] \wl5528 and [O I] \wl 5577 lines with two Gaussians, with the results reported in Table \ref{table:oilines}. 

At the temperatures of interest here, the critical densities (above which the lines form in LTE) for the [O I] \wl5577 and [O I] \wll6300, 6364 lines are $n_e^{5577} = 10^8 \beta_{5577}$ cm$^{-3}$ and $n_e^{6300,6364} = 3\e{6} \beta_{6300,6364}$ cm$^{-3}$, respectively, where $\beta_\lambda = \left(1-\exp{\left(-\tau_\lambda\right)}\right)/\tau_\lambda$ is the Sobolev escape probability and $\tau_\lambda$ is the Sobolev optical depth \citep{Sobolev1957}.
 The [O I] \wll 6300, 6364 critical density is well below the oxygen-zone electron density in the model for several years, implying that the line is formed in LTE. For the [O I] \wl 5577 line, the LTE approximation is accurate up to $\sim$250 days, but after that the model luminosity starts falling below the LTE value.

In LTE, the line ratio is 
\begin{flalign}
&\frac{L_{5577}}{L_{6300, 6364}} = \frac{g_{2p^4(^1S)}}{g_{2p^4(^1D)}} e^{\frac{-\Delta E}{kT}}\frac{A_{5577}\beta_{5577}}{A_{6300,6364} \beta_{6300,6364}}\frac{\lambda_{6300,6364}}{\lambda_{5577}} \nonumber \\
&= 38 \times \exp{\left(\frac{-25790\ K}{T}\right)}\frac{\beta_{5577}}{\beta_{6300, 6364}}~. 
\label{eq:ratio}
\end{flalign}
where we have used $\Delta E = E(2p^4(^1S))-E(2p^{4}(^1D)) = 2.22\ \mbox{eV}$, $g(2p^4(^1S))=1$, $g(2p^{4}(^1D))=5$, $A_{6300,6364}=7.4\e{-3}$ s$^{-1}$, and $A_{5577}=1.26$ s$^{-1}$.
If the [O I] \wl 5577 line is formed in NLTE, the line ratio will be smaller than the quantity on the RHS, still allowing a \emph{lower limit} to $T$ to be set. This in turn translates to an \emph{upper limit} in the oxygen mass, which we will find useful. 

A direct and model-independent determination of $T$ is complicated by the factor $\beta_{5577}/\beta_{6300,6364}$, but fortunately this can be constrained. Since the observed [O I] \wl 6364 line is always weaker than [O I] \wl 6300, the doublet is at least partially transitioning to optical thinness (in the optically thick limit the two lines have equal strength whereas in the optically thin limit the [O I] \wl6364 line is three times weaker than the [O I] \wl6300 line), so $\tau_{6300,6364} \lesssim 2$, or equivalently $\beta_{6300,6364} \gtrsim 0.5$.
The optical depth in [O I] \wl 5577 is always smaller than in [O I] \wll 6300, 6364 (at least for the $T \lesssim 5000$ K regime that is relevant here), so also $\beta_{5577} \gtrsim 0.5$. Thus, the $\beta_{5577}/\beta_{6300,6364}$ ratio must be in the range $1-2$.
In our models, the ratio is between $1.3 - 1.6$ over the evolution covered here ($250 - 450$ days). If we use a value of 1.5 throughout, the observed line ratios correspond to the temperatures listed in column five of Table \ref{table:oilines}. 


These temperatures are $\sim$500 K lower than the ones computed in the model. Part of this may be due to NLTE in [O I] \wl 5577, which causes the temperatures computed assuming LTE to be lower than the true temperatures. Inspection of our statistical equilibrium solutions indeed shows that the parent state of [O I] \wl 5577 (2p$^4$($^1$S)) deviates from LTE after 250 days, with a departure coefficient $n_{2p^4(1S)}/n_{2p^4(1S)}^{LTE}=0.8-0.3$ over the $250-450$ day interval. Nevertheless, the [O I] \wl 5577 line in the model is somewhat brighter than the observed line (Fig. \ref{fig:oi5577}), so part of the temperature discrepancy may also be due to a too high gamma-ray deposition into the oxygen clumps in the model, in turn caused by a too strong mixing between $^{56}$Ni and oxygen clumps. Although the mixing affects the thermal conditions in the oxygen zones, its influence via the ionization balance is small as almost all oxygen is neutral.

Given a temperature estimate, the O I mass can be estimated from the luminosity in either line. The weaker exponential dependency of the [O I] \wll 6300, 6364 luminosity with temperature propagates a smaller error from temperature errors, and in addition $L_{6300,6364}$ can be measured to higher accuracy than $L_{5577}$ since it is always stronger. We thus use the equation
\begin{align}
M_{\rm OI} =  L_{6300,6364} 16 m_p \frac{Z(T)}{g_{2p^4(^1D)}}e^{\frac{E_{2p^4(^1D)}}{kT}} \left(A \beta  h \nu\right)^{-1}_{6300,6364} \nonumber \\
= \frac{L_{6300,6364} / \beta_{6300, 6364} }{9.7\e{41}\ \mbox{erg s}^{-1}}\times \exp{\left(\frac{22720\ K}{T}\right)}\ M_\odot\ ,
\label{eq:oimasses}
\end{align}
to estimate the oxygen mass (where $m_p$ is the proton mass, and we approximate the partition function $Z(T)$ with the ground state statistical weight, $g_{2p^4(^3P)}=9$). If we use $\beta_{6300,6364}=0.5$ throughout (its minimum value based on the [O I] \wl 6300/[O I] \wl 6364 line ratio)
, we obtain the O I masses listed in the last column of Table \ref{table:oilines} ($\sim0.6$ \msun). These are close to the total masses of thermally emitting oxygen as the gas is mostly neutral ($x_{\rm OII} \lesssim 0.1$ at all times in the models). 
The derived masses are in the range $0.4-0.9$ \msun\ at all epochs, which corresponds to the amount of oxygen in the O/Ne/Mg layer of a $16-17$ \msun\ progenitor as computed by \citet{Woosley2007}. Using a $\beta_{5577}/\beta_{6300,6364}$ ratio of 2 instead of 1.5 shifts the oxygen mass range to $0.6-1.2$ \msun\ ($M_{\rm ZAMS}=17-19\ M_\odot$), and using a ratio of 1 gives $0.3-0.7$ \msun\ ($M_{\rm ZAMS}=14-17\ M_\odot$). The oxygen mass in the O/Ne/Mg layer of the 15 \msun\ progenitor star is 0.3 \msun, on the lower end of these estimates, for the reasons discussed above (some combination of departure from LTE and a too strong mixing between $^{56}$Ni and oxygen clumps in the model).

The power of these  analytical arguments is that the [O I] \wl 6300/[O I] \wl 6364 line ratio combined with the [O I] \wl 5577/[O I] \wll 6300,~6364 line ratio provide checks on the density and energy deposition for the oxygen clumps in any model. Since breakdown of the LTE assumption can only lead to overestimated oxygen masses using Eqs. \ref{eq:ratio} and \ref{eq:oimasses}, it  would be difficult to reconcile any ejecta which has more than 1 \msun\ of oxygen in the O/Ne/Mg zone with the observed oxygen line strengths, \emph{independent} of how the oxygen is mixed hydrodynamically and compositionally. For example,  at $M_{\rm ZAMS}= 20$ \msun, the oxygen mass in the O/Ne/Mg layer is over 1.5 \msun \ in the \citet{Woosley2007} models, and the spectrum from such an ejecta would be inconsistent with the observed line strengths of SN 2012aw given the constraints on temperature derived here (see J12 for model spectra from such a high-mass star). It also seems clear that a hypothetical inmixing of coolants like calcium, carbon, silicon, CO, or SiO into the O/Ne/Mg zone cannot save such a model; apart from the temperature constraints considered above, Figs. \ref{fig:optical} and \ref{fig:NIR} show that there is not sufficient observed luminosity from either of these elements ([Ca II] \wll 7291, 7323, Ca II \wl \wl 8498, 8542, 8662, [C I] \wl8727, [C I] \wll9824, 9850, [Si I] \wl1.10 $\upmu$m, [Si I] \wll 1.60, 1.64 $\upmu$m, CO overtone band), to account for any significant cooling of a large oxygen mass. For SN 2004et, which had a very similar spectral evolution in the optical and NIR, observations in the MIR also ruled out CO and SiO fundamental bands as strong enough cooling channels for such a scenario.

The models here therefore represent calculations with no obvious discrepancy with observations regarding the thermal emission from the oxygen zones.
We find, both here for SN 2012aw and in J12 for SN 2004et, an amount of thermally emitting oxygen $<1$ \msun. As the oxygen mass rises steeply with progenitor mass, this sets a constraint on the upper mass for the progenitor star model of 18 \msun, assuming nucleosynthesis as calculated by \citet{Woosley2007}.  

\begin{table*}
\centering
\caption{The line luminosities of [O I] \wl 5577 and [O I] \wll 6300, 6364, their ratio, the single-zone LTE temperature corresponding to this ratio, and the LTE O I mass corresponding to this temperature. For all line luminosity measurements we estimate a $\pm$30\% (relative) error in [O I] \wl5577 and a $\pm$20\% error in [O I] \wll 6300, 6364.}
\begin{tabular}{|c|c|c|c|c|c|}
\hline
Time   & $L_{5577}$              & $L_{6300, 6364}$            & $L_{5577}/L_{6300,6364}$                & $T^{LTE}_{\beta_{ratio}=1.5}$       & $M(O I)^{LTE}_{\beta_{ratio}=1.5,\beta_{6300,6364}=0.5}$\\
(days) & ($10^{38}$ erg s$^{-1}$)          & ($10^{38}$ erg s$^{-1}$)            &                    & (K)        & (\msun)\\
\hline
250 & $1.5\pm 0.45$     & $13 \pm{2.6}$    &  $0.12 \pm 0.042$  & $4170_{-280}^{+220}$  & $0.6_{-0.2}^{+0.3}$\\
332 & $0.68\pm 0.20$    & $8.6 \pm 1.7$    & $0.079 \pm 0.029$  & $3920_{-250}^{+190}$  & $0.6_{-0.1}^{+0.3}$\\
369 & $0.39\pm 0.12$    & $6.7 \pm 1.4$   & $0.057 \pm 0.021$  & $3740_{-230}^{+170}$  & $0.6_{-0.2}^{+0.3}$ \\
451 & $0.082\pm 0.025$  & $3.3 \pm {0.66}$   & $0.025 \pm 0.0090$ & $3330_{-180}^{+140}$  & $0.6_{-0.1}^{+0.3}$ \\
\hline
\end{tabular}
\label{table:oilines}
\end{table*}


The only other clearly detected oxygen line is O I \wl 1.130 $\upmu$m (O I \wl7771 may be detected in the 250-day spectrum but is not to be confused with the (unidentified)\footnote{A plausible identification of this line is the ground state resonance line of potassium - K I 4s($^2$S)-4p($^2$P$^o$) \wll 7665, 7699 \citep{Spyromilio1991b} which is not included in our model.} line seen at $\sim$7700 \AA). The O I \wl 1.130 $\upmu$m  line is radiatively pumped by Ly$\beta$ line overlap with O I \wl1025.76, much of which is zone-crossing between hydrogen clumps and oxygen clumps\footnote{The Ly$\beta$ photon is created in a hydrogen clump but performs a random walk into an oxygen clump while still resonantly trapped. This leads to a breakdown of the Sobolev approximation where all physical conditions are assumed to stay constant over the resonance region.} \citep[][J12]{Oliva1993}. Our model treats the \emph{local} line-overlap in this line, producing an emission feature from the primordial oxygen in the hydrogen clumps (Fig. \ref{fig:NIR}). As is clear from Fig. \ref{fig:NIR} though, this is not enough to reproduce the line luminosity, in agreement with the analysis of this line in SN 1987A \citep{Oliva1993}. 
That the [O I] \wl 1.130 $\upmu$m line is not a recombination line can be inferred from the lack of similar strength emission from other lines that are expected to have similar recombination luminosities; O I \wl7771, O I \wl9263 and O I \wl 1.316 $\upmu$m. The oxygen recombination lines are all observationally weak or absent. In the model this arises as many O II ions are neutralized by rapid charge transfer reactions, with for example C I and Mg I, rather than by radiative recombination. These charge transfer reactions usually occur to the ground state or first excited state in O I, thereby preventing the radiative cascade that follows radiative recombinations to higher levels.

\begin{figure*}
\includegraphics[trim=1mm 1mm 1mm 1mm, clip, width=0.8\linewidth]{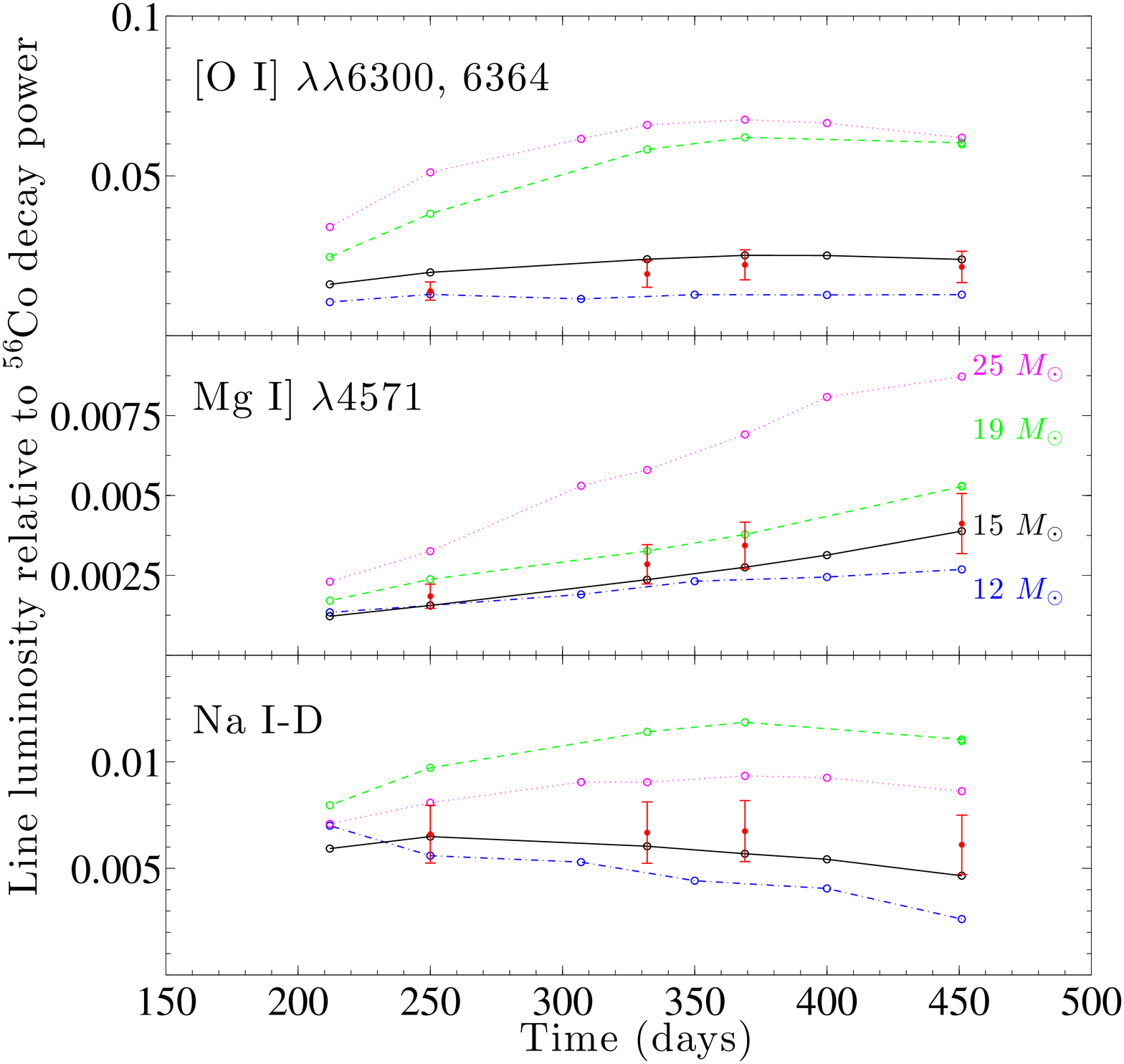}  
\caption{Observed luminosities of [O I] \wll6300, 6364, Mg I] \wl4571, and Na I-D lines compared to models for 12 (blue, dot-dashed), 15 (black, solid), 19 (green, dashed), and 25 \msun\ (magenta, dotted) progenitors. The line fluxes are extracted by an automated algorithm applied in the same way to observed and modeled spectra, as described in J12. We estimate the statistical errors by (quadrature) adding the RMS error of the photometric flux calibration (Sect. \ref{sec:obs}) to the estimated error from the continuum fit in the algorithm (taken as $\pm$20\% here).}
\label{fig:linefluxes}
\end{figure*}

Two other important elements reside in the carbon burning ashes; magnesium and sodium. A successful model must be able to account for the observed luminosities in these lines as well as in the oxygen lines, but as discussed in J12, these lines are more challenging to model accurately. Fig. \ref{fig:linefluxes} shows the measured evolution in the line luminosities of [O I] \wll6300, 6364, Mg I] \wl4571, and Na I-D, compared to the models for 12, 15, 19 and 25 \msun~progenitors.  
As seen in Fig. \ref{fig:linefluxes}, the observed Mg I] \wl4571 and Na I-D line strengths
are bracketed by the $15-19$ \msun\ model range, which overlaps with the 14 - 18 \msun \ mass range determined from the oxygen lines. 

The 25 \msun\ model shows some interesting behaviour. It has an O/Ne/Mg zone of 2.7 \msun, compared to 1.9 \msun\ of the 19 \msun\ model and 0.45 \msun\ of the 15 \msun\ model. There is thus a more dramatic increase in oxygen production between 15 and 19 \msun\ than between 19 and 25 \msun, which is reflected in the oxygen line luminosities. The 25 \msun\ model still produces brighter O I lines, but only by $\sim$20\% between $140-350$ days and almost not at all after that. In addition to the smaller difference in oxygen production, the metal content in the core of the 25 \msun\ ejecta starts to take up such a large fraction of the volume (50\%), that some line blocking occurs even around 6300 \AA, further reducing the line luminosity. 
The Mg I] 4571 line luminosity increases more linearly throughout the $12-25$ \msun\ range, and is therefore an important complement to the oxygen lines for the analysis. Finally, the Na I-D lines are actually weaker in the 25 \msun\ model compared to the 19 \msun\ model. This is because sodium nucleosynthesis is not a strictly monotonic function of progenitor mass, and there is less synthesized sodium in the 25 \msun\ ejecta ($6\e{-3}$ \msun) than in the 19 \msun\ ejecta ($1.3\e{-2}$ \msun).

In the NIR, the model reproduces all the distinct emission lines qualitatively, which includes lines from H I, He I, C I, N I, O I, Na I, Mg I, Si I, S II, Fe I, Fe II, Co II, as labeled in Fig. \ref{fig:NIR}. The model predicts a strong He I \wl2.058 $\upmu$m line which is not observed (it was also absent/weak in SN 2004et), as well as a too strong He I \wl 1.083 $\upmu$m line. Of the NIR lines, J12 found C I \wl1.454 $\upmu$m,  Mg I \wl1.504 $\upmu$m and [Si I] \wll1.607, 1.645 $\upmu$m to be the lines most sensitive to the nucleosynthesis. In general these lines are in rough agreement with the 15 \msun~model (Fig. \ref{fig:NIR}), but we cannot with confidence distinguish between different models in the $12-19$ \msun\ range. For instance the [Si I] \wll 1.607, 1.645 $\upmu$m lines are in best agreement with a 12 \msun\ model whereas Mg I \wl 1.504 $\upmu$m is in best agreement with a 19 \msun\ model. The uncertainty in the absolute flux calibration of the NIR spectrum (as NIR photometry is lacking) should also be kept in mind.  
All observed line fluxes appear weaker than in the 19 \msun\ model (see J12 for the 19 \msun\ spectrum), and the NIR spectrum constraints therefore agree with the optical constraints in ruling out a high-mass progenitor.

\begin{figure}
\includegraphics[trim=16mm 3mm 8mm 10mm, clip, width=1\linewidth]{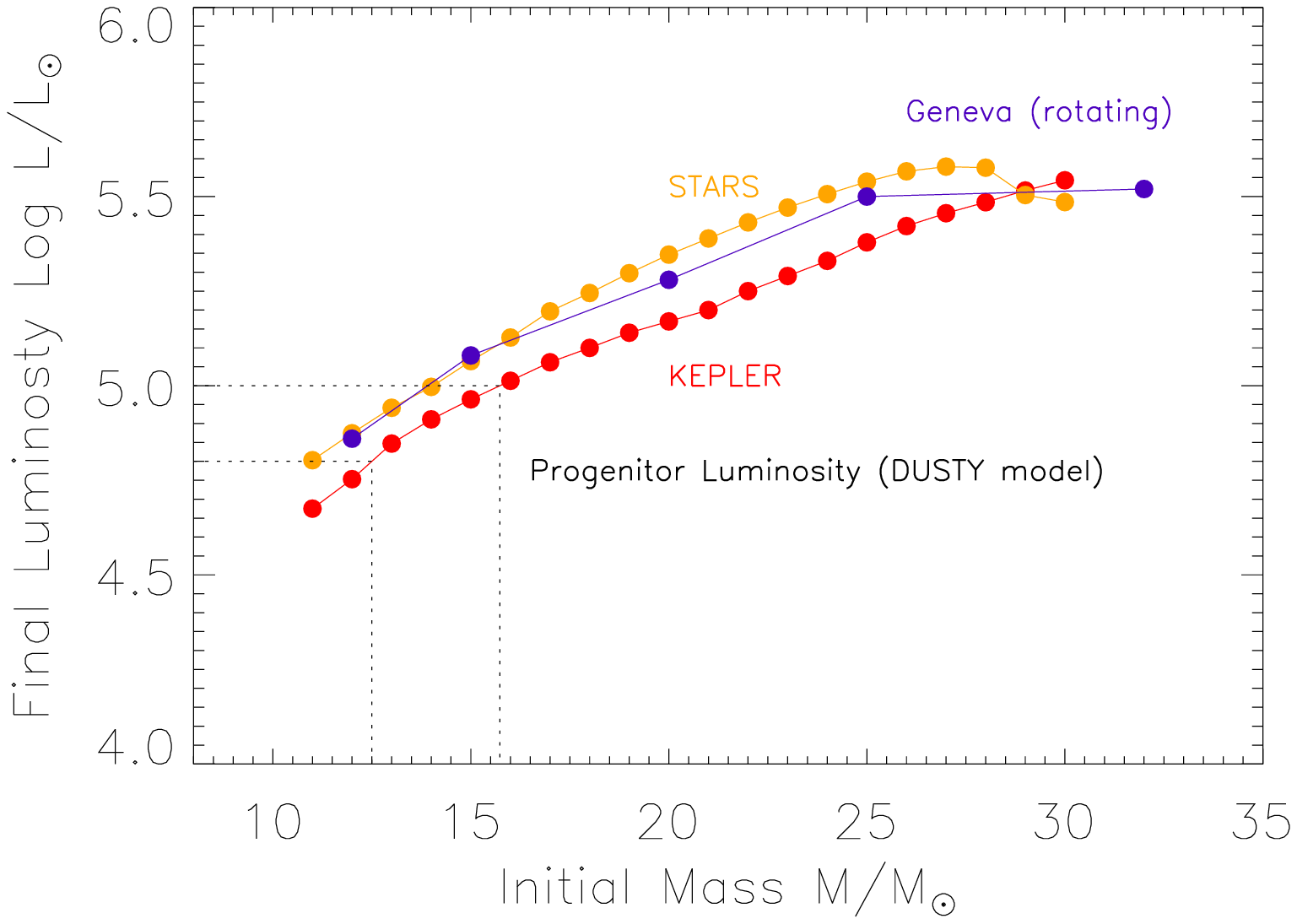}
\caption{The final pre-SN luminosity as function of stellar mass, from three sets of stellar evolution models discussed in the text. The KEPLER models are those used as input for the spectral modelling in this paper. The range of luminosities of the progenitor star of SN 2012aw calculated by \citet*{Kochanek2012} ($\log{L/L_\odot} = 4.8-5.0$) are shown, corresponding to the end point luminosities of $13-16$ \msun\ KEPLER progenitors.}
\label{fig:endpoints}
\end{figure}

\section{Discussion}
\label{sec:discussion}
The link between progenitor mass and nucleosynthesis depends on how some uncertain physical processes are treated in the stellar evolution model. Whereas mass-loss, convection, and metallicity have weak influence on the oxygen nucleosynthesis \citep{Dessart2013}, semi-convection \citep{Langer1991}, overshooting \citep{Langer1991, Schaller1992, Dessart2013} and rotation \citep{Hirschi2004, Dessart2013} can alter the oxygen nucleosynthesis by more than a factor of two. The KEPLER models we use employ efficient semi-convection \citep[see discussions in][]{Langer1985, Woosley1988, Langer1991} and overshooting 
\citep[see][]{Woosley1988}, but no rotation. 
Models with overshooting produce about twice as much oxygen as models without and fast-rotating models produce a factor $\sim$1.5 more oxygen than non-rotating ones \citep{Dessart2013}. 
These effects are also reflected in the stellar luminosities, with non-rotating models with efficient overshooting 
tending to match the luminosities of rotating models
\citep[see discussion in][]{Smartt2009b}. 



Reviewing a variety of models for nucleosynthesis yields \citep{Thielemann1996, Nomoto1997, Limongi2003, Hirschi2005, Woosley2007}, the factor two variety in oxygen yield for a given ZAMS mass is confirmed. For instance, at $M_{\rm ZAMS}=15$ \msun, the ejected masses vary from $0.4-1.0$ \msun, and at $M_{\rm ZAMS}=25$ \msun from $2.2-3.6$ \msun. This latter range is clearly higher than the oxygen mass in SN 2012aw derived here.


Whereas these issues in stellar evolution physics still remain to be resolved, any model can be checked for its ability to reproduce \emph{both} the progenitor appearance and the derived nucleosynthesis of an individual event. As described in Sect. 3, the KEPLER supernova progenitor models with masses in the range $M_{\rm ZAMS} = 14-18$ \msun\ have nucleosynthesis that satisfactorily reproduces the observed nebular spectrum of SN 2012aw. The model progenitor luminosities from this range can then be compared with the derived luminosity of the progenitor star. 
The \citet{Woosley2007} final luminosities (from private communication with S. Woosley) are shown in Fig.\,\ref{fig:endpoints}, along with those 
from the STARS models \citep{Eldridge2004}\footnote{http://www.ast.cam.ac.uk/$\sim$stars/archive/}
and the Geneva rotating models \citep{Ekstrom2012}. The KEPLER model luminosities are at the pre-SN silicon burning stage, whereas the other two are 
at the end of core C-burning, but the surface luminosity of a red supergiant is not expected to change after this stage.

The progenitor star of SN 2012aw was originally identified from optical and near-infrared images by \cite{Fraser2012} and \cite{vandyk2012} as a red supergiant with luminosity of  $\log  \left(L/L_\odot\right) = 5.0 - 5.6$  and $\log \left(L/L_\odot\right) = 5.18-5.24$,  respectively. Both papers argued for significant extinction of the progenitor, most likely due to circumstellar dust that was later destroyed in the explosion, as the SN itself had a low  line of sight extinction. An extended reanalysis of the progenitor photometry was carried out by 
\citet{Kochanek2012} who performed a more detailed calculation of the extinction by computing the transfer through a realistic circumstellar density distribution and computing optical properties of the dust (in comparison, \citet{Fraser2012} and \citet{vandyk2012} treated the dust as a distant dust screen which removes all photons that scatter on it from the observer). 
With the \citet{Kochanek2012} treatment, the derived stellar luminosity is lowered by several tenths of a dex, as many photons scatter on the circumstellar dust  but still reach the observer, and the authors find the observed  progenitor SED to be consistent with a red supergiant star with 
$\log \left(L/L_\odot\right) = 4.8-5.0$. As shown in Fig. \ref{fig:endpoints}, this progenitor luminosity corresponds to a KEPLER progenitor in the $13-16$ \msun\ range, consistent with the results from the nebular analysis in this paper. 

 \citet{Kochanek2012} also noted that the X-ray and radio observations of SN 2012aw constrained the progenitor's mass-loss rate to $-\log \left({\dot{M}/M_\odot \mbox{yr}^{-1}}\right) = 5.5 - 5.0$. The KEPLER models employ mass loss rates from the empirical parametrisation of  \cite{Nieu1990}
which are $-\log \left({\dot{M}/M_\odot \mbox{yr}^{-1}}\right) = 5.4-5.2$ for $\log \left(L/L_\odot\right) = 5.0$, $M_{\rm ZAMS} = 15$ \msun\ and $3500\ \mbox{K} < T_{\rm eff} < 4500\ \mbox{K}$. 

There is therefore encouraging consistency between the required mass-loss rate from X-ray and radio observations, the progenitor luminosity from pre-explosion imaging, and the nucleosynthesis derived from the nebular spectra. However, hydrodynamical modeling of the optically thick phase \citep{dallOra2013} favors an ejecta mass $>20$ \msun, and thus a higher-mass progenitor. Understanding the differences in results between progenitor imaging, hydrodynamical modeling, and nebular phase spectral analysis is a high priority in the Type IIP research field.

We hope in the future to continually increase the accuracy of our nebular-phase spectral models, and apply them to grids of multidimensional explosions with self-consistently calculated mixing. Combined with further high signal-to-noise optical and near-infrared nebular phase spectra of nearby SNe, we will eventually be able to constrain the yields of individual elements to still higher accuracy and improve our understanding of stellar and supernova nucleosynthesis.

\section{Conclusions}
We have obtained optical and near-infrared spectra of the Type IIP SN 2012aw in the nebular phase, and analyzed these with spectral model calculations. 
The observed spectral evolution shows close agreement with a  model for a $M_{\rm ZAMS}= 15\ M_\odot$ progenitor (evolved and exploded with KEPLER), with additional analysis of magnesium and sodium emission lines consistently pointing to a progenitor in the $14-18$ \msun\ range. The endpoint luminosities of $14-16$ \msun\ models are also consistent with the estimate of the luminosity of the directly detected progenitor star by \citet{Kochanek2012}, and hence a self-consistent solution is found.
 We have demonstrated how combining the individual evolution in the three oxygen lines excited by thermal collisions ([O I] \wl5577, [O I] \wl 6300, and [O I] \wl6364) can be used to constrain the mass of the oxygen present in the carbon burning ashes (here limiting it to $<$ 1 \msun), independent of uncertainties in the hydrodynamical and chemical mixing. 

Reviewing the literature of published nebular spectra of Type IIP SNe (with ``normal'' explosion energies and $^{56}$Ni masses)\footnote{Some objects in the ``subluminous'' class of Type IIP SNe \citep[e.g.][]{Turatto1998, Benetti2001, Pastorello2004, Fraser2011}, which have extremely low kinetic energies and $^{56}$Ni masses, may still be associated with such massive stars (although those with directly identified progenitors so far have $M_{ZAMS}<12$ \msun), but a viewpoint on this from their nebular spectra would need models calculations with significantly different parameters than those used here.} \citep{Turatto1993, Schmidt1993, Benetti1994, Elmhamdi2003, Pozzo2006, Sahu2006, Quimby2007, Maguire2010, Andrews2011, Inserra2011, Roy2011, Tomasella2013},  we find no observations where the [O I] \wll 6300, 6364 lines are significantly stronger (relative to the optical spectrum as a whole) than in SN 2012aw. We conclude that no Type IIP SN has yet been shown to eject nucleosynthesis products expected from stars more massive than 20 \msun. 


\section*{Acknowledgments}
The research leading to these results has received funding from the European Research Council under the European Union's Seventh Framework Programme (FP7/2007-2013)/ERC Grant agreement n$^{\rm o}$ [291222]  (PI : S. J. Smartt).
The results are based on observations made with the William Herschel Telescope (PATT and Service time, proposal ID SW2012b27) and the Nordic Optical Telescope operated on the island of La Palma by the Isaac Newton Group in the Spanish Observatorio del Roque de los Muchachos of the Instituto de Astrof\'{i}sica de Canarias.
We thank Cosimo Inserra, Ting-Wan Chen, and Elisabeth Gall for discussion and observing assistance, and  Andy Lawrence, Alex Bruce, Massimo Dall'Ora, Stan Woosley, and Christopher Kochanek for discussion.

\bibliographystyle{mn2e3}
\bibliography{bibl}

\begin{thebibliography}{62}
\expandafter\ifx\csname natexlab\endcsname\relax\def\natexlab#1{#1}\fi

\bibitem[{{Andrews} {et~al}\mbox{.}(2011){Andrews}, {Sugerman}, {Clayton},
  {Gallagher}, {Barlow}, {Clem}, {Ercolano}, {Fabbri}, {Meixner}, {Otsuka},
  {Welch}, \& {Wesson}}]{Andrews2011}
{Andrews} J.~E. {et~al.}, 2011, \apj, 731, 47

\bibitem[{{Benetti} {et~al}\mbox{.}(1994){Benetti}, {Cappellaro}, {Turatto},
  {della Valle}, {Mazzali}, \& {Gouiffes}}]{Benetti1994}
{Benetti} S., {Cappellaro} E., {Turatto} M., {della Valle} M., {Mazzali} P.~A.,
  {Gouiffes} C., 1994, \aap, 285, 147

\bibitem[{{Benetti} {et~al}\mbox{.}(2001){Benetti}, {Turatto}, {Balberg},
  {Zampieri}, {Shapiro}, {Cappellaro}, {Nomoto}, {Nakamura}, {Mazzali}, \&
  {Patat}}]{Benetti2001}
{Benetti} S. {et~al.}, 2001, \mnras, 322, 361

\bibitem[{{Bersten}, {Benvenuto} \& {Hamuy}(2011){Bersten}, {Benvenuto}, \&
  {Hamuy}}]{Bersten2011}
{Bersten} M.~C., {Benvenuto} O., {Hamuy} M., 2011, \apj, 729, 61

\bibitem[{{Bose} {et~al}\mbox{.}(2013){Bose}, {Kumar}, {Sutaria}, {Kumar},
  {Roy}, {Bhatt}, {Pandey}, {Chandola}, {Sagar}, {Misra}, \&
  {Chakraborti}}]{Bose2013}
{Bose} S. {et~al.}, 2013, \mnras, 433, 1871

\bibitem[{{Cardelli}, {Clayton} \& {Mathis}(1989){Cardelli}, {Clayton}, \&
  {Mathis}}]{Cardelli1989}
{Cardelli} J.~A., {Clayton} G.~C., {Mathis} J.~S., 1989, \apj, 345, 245

\bibitem[{{Dall'Ora} {et~al}\mbox{.}(2013){Dall'Ora}, {Botticella}, {Pumo},
  {Zampieri}, {Tomasella}, {Pignata}, {Bayless}, {Pritchard}, {Kotak}, \&
  {Inserra}}]{dallOra2013}
{Dall'Ora} M. {et~al.}, 2013, \apj, submitted

\bibitem[{{Dessart} \& {Hillier}(2011)}]{Dessart2011}
{Dessart} L., {Hillier} D.~J., 2011, \mnras, 410, 1739

\bibitem[{{Dessart} {et~al}\mbox{.}(2013){Dessart}, {Hillier}, {Waldman}, \&
  {Livne}}]{Dessart2013}
{Dessart} L., {Hillier} D.~J., {Waldman} R., {Livne} E., 2013, \mnras, 433,
  1745

\bibitem[{{Ekstr{\"o}m} {et~al}\mbox{.}(2012){Ekstr{\"o}m}, {Georgy},
  {Eggenberger}, {Meynet}, {Mowlavi}, {Wyttenbach}, {Granada}, {Decressin},
  {Hirschi}, {Frischknecht}, {Charbonnel}, \& {Maeder}}]{Ekstrom2012}
{Ekstr{\"o}m} S. {et~al.}, 2012, \aap, 537, A146

\bibitem[{{Eldridge} \& {Tout}(2004)}]{Eldridge2004}
{Eldridge} J.~J., {Tout} C.~A., 2004, \mnras, 353, 87

\bibitem[{{Elmhamdi} {et~al}\mbox{.}(2003){Elmhamdi}, {Danziger}, {Chugai},
  {Pastorello}, {Turatto}, {Cappellaro}, {Altavilla}, {Benetti}, {Patat}, \&
  {Salvo}}]{Elmhamdi2003}
{Elmhamdi} A. {et~al.}, 2003, \mnras, 338, 939

\bibitem[{{Fransson} \& {Chevalier}(1989)}]{Fransson1989}
{Fransson} C., {Chevalier} R.~A., 1989, \apj, 343, 323

\bibitem[{{Fraser} {et~al}\mbox{.}(2011){Fraser}, {Ergon}, {Eldridge},
  {Valenti}, {Pastorello}, {Sollerman}, {Smartt}, {Agnoletto}, {Arcavi},
  {Benetti}, {Botticella}, {Bufano}, {Campillay}, {Crockett}, {Gal-Yam},
  {Kankare}, {Leloudas}, {Maguire}, {Mattila}, {Maund}, {Salgado}, {Stephens},
  {Taubenberger}, \& {Turatto}}]{Fraser2011}
{Fraser} M. {et~al.}, 2011, \mnras, 417, 1417

\bibitem[{{Fraser} {et~al}\mbox{.}(2012){Fraser}, {Maund}, {Smartt},
  {Botticella}, {Dall'Ora}, {Inserra}, {Tomasella}, {Benetti}, {Ciroi},
  {Eldridge}, {Ergon}, {Kotak}, {Mattila}, {Ochner}, {Pastorello}, {Reilly},
  {Sollerman}, {Stephens}, {Taddia}, \& {Valenti}}]{Fraser2012}
{Fraser} M. {et~al.}, 2012, \apjl, 759, L13

\bibitem[{{Hirschi}, {Meynet} \& {Maeder}(2004){Hirschi}, {Meynet}, \&
  {Maeder}}]{Hirschi2004}
{Hirschi} R., {Meynet} G., {Maeder} A., 2004, \aap, 425, 649

\bibitem[{{Hirschi}, {Meynet} \& {Maeder}(2005){Hirschi}, {Meynet}, \&
  {Maeder}}]{Hirschi2005}
{Hirschi} R., {Meynet} G., {Maeder} A., 2005, \aap, 433, 1013

\bibitem[{{Inserra} {et~al}\mbox{.}(2011){Inserra}, {Turatto}, {Pastorello},
  {Benetti}, {Cappellaro}, {Pumo}, {Zampieri}, {Agnoletto}, {Bufano},
  {Botticella}, {Della Valle}, {Elias Rosa}, {Iijima}, {Spiro}, \&
  {Valenti}}]{Inserra2011}
{Inserra} C. {et~al.}, 2011, \mnras, 417, 261

\bibitem[{{Jerkstrand}, {Fransson} \& {Kozma}(2011){Jerkstrand}, {Fransson}, \&
  {Kozma}}]{Jerkstrand2011}
{Jerkstrand} A., {Fransson} C., {Kozma} C., 2011, \aap, 530, A45

\bibitem[{{Jerkstrand} {et~al}\mbox{.}(2012){Jerkstrand}, {Fransson},
  {Maguire}, {Smartt}, {Ergon}, \& {Spyromilio}}]{Jerkstrand2012}
{Jerkstrand} A., {Fransson} C., {Maguire} K., {Smartt} S., {Ergon} M.,
  {Spyromilio} J., 2012, \aap, 546, A28, (J12)

\bibitem[{{Kifonidis} {et~al}\mbox{.}(2006){Kifonidis}, {Plewa}, {Scheck},
  {Janka}, \& {M{\"u}ller}}]{Kifonidis2006}
{Kifonidis} K., {Plewa} T., {Scheck} L., {Janka} H.-T., {M{\"u}ller} E., 2006,
  \aap, 453, 661

\bibitem[{{Kochanek}, {Khan} \& {Dai}(2012){Kochanek}, {Khan}, \&
  {Dai}}]{Kochanek2012}
{Kochanek} C.~S., {Khan} R., {Dai} X., 2012, \apj, 759, 20

\bibitem[{{Kotak} {et~al}\mbox{.}(2009){Kotak}, {Meikle}, {Farrah}, {Gerardy},
  {Foley}, {Van Dyk}, {Fransson}, {Lundqvist}, {Sollerman}, {Fesen},
  {Filippenko}, {Mattila}, {Silverman}, {Andersen}, {H{\"o}flich}, {Pozzo}, \&
  {Wheeler}}]{Kotak2009}
{Kotak} R. {et~al.}, 2009, \apj, 704, 306

\bibitem[{{Langer}(1991)}]{Langer1991}
{Langer} N., 1991, \aap, 252, 669

\bibitem[{{Langer}, {El Eid} \& {Fricke}(1985){Langer}, {El Eid}, \&
  {Fricke}}]{Langer1985}
{Langer} N., {El Eid} M.~F., {Fricke} K.~J., 1985, \aap, 145, 179

\bibitem[{{Li} \& {McCray}(1992)}]{Li1992}
{Li} H., {McCray} R., 1992, \apj, 387, 309

\bibitem[{{Li} {et~al}\mbox{.}(2011){Li}, {Leaman}, {Chornock}, {Filippenko},
  {Poznanski}, {Ganeshalingam}, {Wang}, {Modjaz}, {Jha}, {Foley}, \&
  {Smith}}]{Li2011}
{Li} W. {et~al.}, 2011, \mnras, 412, 1441

\bibitem[{{Limongi} \& {Chieffi}(2003)}]{Limongi2003}
{Limongi} M., {Chieffi} A., 2003, \apj, 592, 404

\bibitem[{{Liu} \& {Dalgarno}(1995)}]{Liu1995}
{Liu} W., {Dalgarno} A., 1995, \apj, 454, 472

\bibitem[{{Maguire} {et~al}\mbox{.}(2010){Maguire}, {Di Carlo}, {Smartt},
  {Pastorello}, {Tsvetkov}, {Benetti}, {Spiro}, {Arkharov}, {Beccari},
  {Botticella}, {Cappellaro}, {Cristallo}, {Dolci}, {Elias-Rosa}, {Fiaschi},
  {Gorshanov}, {Harutyunyan}, {Larionov}, {Navasardyan}, {Pietrinferni},
  {Raimondo}, {di Rico}, {Valenti}, {Valentini}, \& {Zampieri}}]{Maguire2010}
{Maguire} K. {et~al.}, 2010, \mnras, 404, 981

\bibitem[{{Maguire} {et~al}\mbox{.}(2012){Maguire}, {Jerkstrand}, {Smartt},
  {Fransson}, {Pastorello}, {Benetti}, {Valenti}, {Bufano}, \&
  {Leloudas}}]{Maguire2012}
{Maguire} K. {et~al.}, 2012, \mnras, 420, 3451

\bibitem[{{Maurer} {et~al}\mbox{.}(2011){Maurer}, {Jerkstrand}, {Mazzali},
  {Taubenberger}, {Hachinger}, {Kromer}, {Sim}, \& {Hillebrandt}}]{Maurer2011}
{Maurer} I., {Jerkstrand} A., {Mazzali} P.~A., {Taubenberger} S., {Hachinger}
  S., {Kromer} M., {Sim} S., {Hillebrandt} W., 2011, \mnras, 418, 1517

\bibitem[{{Nieuwenhuijzen} \& {de Jager}(1990)}]{Nieu1990}
{Nieuwenhuijzen} H., {de Jager} C., 1990, \aap, 231, 134

\bibitem[{{Nomoto} {et~al}\mbox{.}(1997){Nomoto}, {Hashimoto}, {Tsujimoto},
  {Thielemann}, {Kishimoto}, {Kubo}, \& {Nakasato}}]{Nomoto1997}
{Nomoto} K., {Hashimoto} M., {Tsujimoto} T., {Thielemann} F.-K., {Kishimoto}
  N., {Kubo} Y., {Nakasato} N., 1997, Nuclear Physics A, 616, 79

\bibitem[{{Oliva}(1993)}]{Oliva1993}
{Oliva} E., 1993, \aap, 276, 415

\bibitem[{{Pastorello} {et~al}\mbox{.}(2004){Pastorello}, {Zampieri},
  {Turatto}, {Cappellaro}, {Meikle}, {Benetti}, {Branch}, {Baron}, {Patat},
  {Armstrong}, {Altavilla}, {Salvo}, \& {Riello}}]{Pastorello2004}
{Pastorello} A. {et~al.}, 2004, \mnras, 347, 74

\bibitem[{{Pozzo} {et~al}\mbox{.}(2006){Pozzo}, {Meikle}, {Rayner}, {Joseph},
  {Filippenko}, {Foley}, {Li}, {Mattila}, \& {Sollerman}}]{Pozzo2006}
{Pozzo} M. {et~al.}, 2006, \mnras, 368, 1169

\bibitem[{{Pumo} \& {Zampieri}(2011)}]{Pumo2011}
{Pumo} M.~L., {Zampieri} L., 2011, \apj, 741, 41

\bibitem[{{Quimby} {et~al}\mbox{.}(2007){Quimby}, {Wheeler}, {H{\"o}flich},
  {Akerlof}, {Brown}, \& {Rykoff}}]{Quimby2007}
{Quimby} R.~M., {Wheeler} J.~C., {H{\"o}flich} P., {Akerlof} C.~W., {Brown}
  P.~J., {Rykoff} E.~S., 2007, \apj, 666, 1093

\bibitem[{{Roy} {et~al}\mbox{.}(2011){Roy}, {Kumar}, {Moskvitin}, {Benetti},
  {Fatkhullin}, {Kumar}, {Misra}, {Bufano}, {Martin}, {Sokolov}, {Pandey},
  {Chandola}, \& {Sagar}}]{Roy2011}
{Roy} R. {et~al.}, 2011, \mnras, 414, 167

\bibitem[{{Rybicki} \& {Lightman}(1979)}]{Rybicki1979}
{Rybicki} G.~B., {Lightman} A.~P., 1979, {Radiative processes in astrophysics}

\bibitem[{{Sahu} {et~al}\mbox{.}(2006){Sahu}, {Anupama}, {Srividya}, \&
  {Muneer}}]{Sahu2006}
{Sahu} D.~K., {Anupama} G.~C., {Srividya} S., {Muneer} S., 2006, \mnras, 372,
  1315

\bibitem[{{Sarangi} \& {Cherchneff}(2013)}]{Sarangi2013}
{Sarangi} A., {Cherchneff} I., 2013, \apj, 776, 107

\bibitem[{{Schaller} {et~al}\mbox{.}(1992){Schaller}, {Schaerer}, {Meynet}, \&
  {Maeder}}]{Schaller1992}
{Schaller} G., {Schaerer} D., {Meynet} G., {Maeder} A., 1992, \aaps, 96, 269

\bibitem[{{Schmidt} {et~al}\mbox{.}(1993){Schmidt}, {Kirshner}, {Schild},
  {Leibundgut}, {Jeffery}, {Willner}, {Peletier}, {Zabludoff}, {Phillips},
  {Suntzeff}, {Hamuy}, {Wells}, {Smith}, {Baldwin}, {Weller}, {Navarette},
  {Gonzalez}, {Filippenko}, {Shields}, {Steidel}, {Perlmutter}, {Pennypacker},
  {Smith}, {Porter}, {Boroson}, {Stathakis}, {Cannon}, {Peters}, {Horine},
  {Freeman}, {Womble}, {Stone}, {Marschall}, {Phillips}, {Saha}, \&
  {Bond}}]{Schmidt1993}
{Schmidt} B.~P. {et~al.}, 1993, \aj, 105, 2236

\bibitem[{{Smartt}(2009)}]{Smartt2009a}
{Smartt} S.~J., 2009, \araa, 47, 63

\bibitem[{{Smartt} {et~al}\mbox{.}(2009){Smartt}, {Eldridge}, {Crockett}, \&
  {Maund}}]{Smartt2009b}
{Smartt} S.~J., {Eldridge} J.~J., {Crockett} R.~M., {Maund} J.~R., 2009,
  \mnras, 395, 1409

\bibitem[{{Sobolev}(1957)}]{Sobolev1957}
{Sobolev} V.~V., 1957, \sovast, 1, 678

\bibitem[{{Spyromilio} \& {Pinto}(1991)}]{Spyromilio1991}
{Spyromilio} J., {Pinto} P.~A., 1991, in European Southern Observatory
  Conference and Workshop Proceedings, Vol.~37, European Southern Observatory
  Conference and Workshop Proceedings, {Danziger} I.~J., {Kjaer} K., eds., p.
  423

\bibitem[{{Spyromilio} {et~al}\mbox{.}(1991){Spyromilio}, {Stathakis},
  {Cannon}, {Waterman}, {Couch}, \& {Dopita}}]{Spyromilio1991b}
{Spyromilio} J., {Stathakis} R.~A., {Cannon} R.~D., {Waterman} L., {Couch}
  W.~J., {Dopita} M.~A., 1991, \mnras, 248, 465

\bibitem[{{Stanishev}(2007)}]{Stanishev2007}
{Stanishev} V., 2007, Astronomische Nachrichten, 328, 948

\bibitem[{{Thielemann}, {Nomoto} \& {Hashimoto}(1996){Thielemann}, {Nomoto}, \&
  {Hashimoto}}]{Thielemann1996}
{Thielemann} F.-K., {Nomoto} K., {Hashimoto} M.-A., 1996, \apj, 460, 408

\bibitem[{{Tomasella} {et~al}\mbox{.}(2013){Tomasella}, {Cappellaro}, {Fraser},
  {Pumo}, {Pastorello}, {Pignata}, {Benetti}, {Bufano}, {Dennefeld},
  {Harutyunyan}, {Iijima}, {Jerkstrand}, {Kankare}, {Kotak}, {Magill},
  {Nascimbeni}, {Ochner}, {Siviero}, {Smartt}, {Sollerman}, {Stanishev},
  {Taddia}, {Taubenberger}, {Turatto}, {Valenti}, {Wright}, \&
  {Zampieri}}]{Tomasella2013}
{Tomasella} L. {et~al.}, 2013, \mnras, 434, 1636

\bibitem[{{Turatto} {et~al}\mbox{.}(1993){Turatto}, {Cappellaro}, {Benetti}, \&
  {Danziger}}]{Turatto1993}
{Turatto} M., {Cappellaro} E., {Benetti} S., {Danziger} I.~J., 1993, \mnras,
  265, 471

\bibitem[{{Turatto} {et~al}\mbox{.}(1998){Turatto}, {Mazzali}, {Young},
  {Nomoto}, {Iwamoto}, {Benetti}, {Cappellaro}, {Danziger}, {de Mello},
  {Phillips}, {Suntzeff}, {Clocchiatti}, {Piemonte}, {Leibundgut},
  {Covarrubias}, {Maza}, \& {Sollerman}}]{Turatto1998}
{Turatto} M. {et~al.}, 1998, \apjl, 498, L129

\bibitem[{{Utrobin} \& {Chugai}(2008)}]{Utrobin2008}
{Utrobin} V.~P., {Chugai} N.~N., 2008, \aap, 491, 507

\bibitem[{{Utrobin} \& {Chugai}(2009)}]{Utrobin2009}
{Utrobin} V.~P., {Chugai} N.~N., 2009, \aap, 506, 829

\bibitem[{{van Dokkum}(2001)}]{vandokkum2001}
{van Dokkum} P.~G., 2001, \pasp, 113, 1420

\bibitem[{{Van Dyk} {et~al}\mbox{.}(2012){Van Dyk}, {Cenko}, {Poznanski},
  {Arcavi}, {Gal-Yam}, {Filippenko}, {Silverio}, {Stockton}, {Cuillandre},
  {Marcy}, {Howard}, \& {Isaacson}}]{vandyk2012}
{Van Dyk} S.~D. {et~al.}, 2012, \apj, 756, 131

\bibitem[{{Woosley} \& {Heger}(2007)}]{Woosley2007}
{Woosley} S.~E., {Heger} A., 2007, \physrep, 442, 269

\bibitem[{{Woosley} \& {Weaver}(1988)}]{Woosley1988}
{Woosley} S.~E., {Weaver} T.~A., 1988, \physrep, 163, 79

\bibitem[{{Woosley} \& {Weaver}(1995)}]{Woosley1995}
{Woosley} S.~E., {Weaver} T.~A., 1995, \apjs, 101, 181

\end{thebibliography}

\label{lastpage}
\end{document}